# Item response parameter estimation performance using Gaussian quadrature and Laplace

Authors: Leticia Arrington[1,2] and Sebastian Ueckert[2]

[1] Department of Pharmacy, Uppsala University, P.O. Box 580, 751 24 Uppsala, Sweden
[2] Corresponding author:  email  leticia.arrington@farmaci.uu.se

## Abstract

Item parameter estimation in pharmacometric item response theory (IRT) models is predominantly performed using the Laplace estimation algorithm as implemented in NONMEM. In psychometrics a wide range of different software tools, including several packages for the open-source software R for implementation of IRT are also available. Each have their own set of benefits and limitations and to date a systematic comparison of the primary estimation algorithms has not been evaluated. A simulation study evaluating varying number of hypothetical sample sizes and item scenarios at baseline was performed using both Laplace and Gauss-hermite quadrature (GHQ-EM). In scenarios with at least 20 items and more than 100 subjects, item parameters were estimated with good precision and were similar between estimation algorithms as demonstrated by several measures of bias and precision. The minimal differences observed for certain parameters or sample size scenarios were reduced when translating to the total score scale. The ease of use, speed of estimation and relative accuracy of the GHQ-EM method employed in *mirt* make it an appropriate alternative or supportive analytical approach to NONMEM for potential pharmacometrics IRT applications.

# Introduction

Clinical assessments that utilize rating scales are often evaluated by summing scores from multiple questions into a single score to assess an individual's ability or disability. Item response theory (IRT) is a statistical methodology for the analysis of these types of composite scores which originated in the field of psychometrics. The methodology describes the relationship between a subject-specific latent variable and the probability of a response for each item through item-specific functions. Some of the strengths of the IRT approach include: (i) a separation of test and subject characteristics, (ii) the transformation of ordinal scale item-level data to interval scale , and (iii) an implicit weighting of the item level data based on their information content. These and other advantages are increasingly appreciated in pharmacometric applications of IRT, as witnessed by a growing number of publications. A particular feature of IRT models is the large number of parameters utilized to parametrize the item-specific or item characteristic functions (ICCs) which need to be estimated from the data. The estimation of these parameters is the focus of this paper.

IRT models are part of the much larger class of nonlinear mixed effect (NLME) models, which are frequently used in pharmacometrics. NLME models describe data in a population accounting for both between subject and within subject variability. NONMEM was one of the first software for NLME modeling and continues to be popular for population pharmacokinetic and pharmacodynamic analyses. It has also been the main modeling tool for pharmacometrics IRT analyses [1].

The popularity of IRT in psychometrics has led to the development of a wide range of different software tools, including several packages for the open- source software R (e.g., *ltm, MCMCpack, eRM and mirt) [2,3,4,5]*. These packages were developed with a focus on psychometrics with limited flexibility for modeling longitudinal data. Nonetheless, their specialized algorithms might be well suited to obtain item parameter estimates for a subsequent longitudinal analysis and hence warrants a closer evaluation. Among the different package options, *mirt* (multidimensional item response theory) has one of the largest sets of features and is actively maintained [5].

Estimation algorithms like Laplace and expectation maximization (EM) are well-suited to handle categorical or discrete ordinal data [1,6]. The challenge in the estimation of parameters for IRT models, and NLME model in general, are the unobserved latent variables (or random effects in the general setting). Multiple algorithms have been proposed to overcome these challenges. NONMEM and mirt alone, offer a wide range of estimation algorithms with varying strengths and weaknesses. In this investigation, we focus on the Laplace estimation algorithm in NONMEM and the expectation maximization (EM) algorithm with Gauss-Hermite quadrature in *mirt* (GHQ-EM). Our choice is based on the widespread use of the NONMEM Laplace algorithm for pharmacometrics IRT analyses and the status of GHQ-EM as the default in *mirt*.

A parameter estimation algorithm for an IRT model needs methodological approaches for two main tasks: (i) the approximation of the intractable marginal likelihood, and (ii) the optimization of the parameter values to maximize the approximated marginal likelihood. Laplace and GHQ-EM approach both parts in a slightly different manner. Laplace utilizes a second order Taylor series approximation at the mode of the joint density (i.e., select the most likely point) to obtain an approximate but tractable individual marginal likelihood expression [1]. This approximated marginal likelihood is then summed for all subjects in the data and optimized in an iterative manner. Since the joint density depends on both data and item parameters, finding the mode and performing the second order Taylor series approximation must be performed for each subject at every iteration. This can be computationally expensive. The GHQ-EM algorithm, as implemented in *mirt*, exploits the assumed normal distribution of the latent variable values in the population to approximate the individual likelihood using Gauss-Hermite quadrature. Gaussian-Hermite quadrature uses a pre-specified grid of points or quadratures across the distribution of etas and weights to replace the integral by the weighted sum of the data density evaluated at the grid points. The grid points are typically centered around zero, giving more weight to values near zero [7]. An increase in grid points tend to result in more precise estimates however it will also result in increased computation time. The sum of the approximated individual marginal likelihoods is then optimized using the EM algorithm [8,9].

It is helpful to look at the marginal likelihood approximation of Laplace and GHQ-EM as two extremes of the same principle. GHQ-EM uses a fixed, potentially large, number of grid points and weights without any consideration for the data at hand. Laplace, in contrast to that, uses only a single grid point and weight at the mode of conditional density but adjusts both to the data present, which can be thought of as an adaptive quadrature [7].

An illustrative comparison of the marginal likelihood approximation from both algorithms is given in below in Figure 1. In the panel in the top left corner, the ICCs for five arbitrary binary items are shown. The second panel, displays the likelihood as a function of the latent variable for a response pattern of 1,1,1,1,1. Panels 3 and 4 show the joint likelihood, the product of the data likelihood and the population prior, as function of the latent variable together with the Laplace and GHQ-EM approximation, respectively.

The aim of this work is to compare the performance of the Laplace and GHQ-EM algorithms for the estimation of the item parameters in IRT models. Herein we investigate the estimation properties and item parameter recovery, for a set of sample size and assessment length scenarios utilizing unidimensional IRT models.

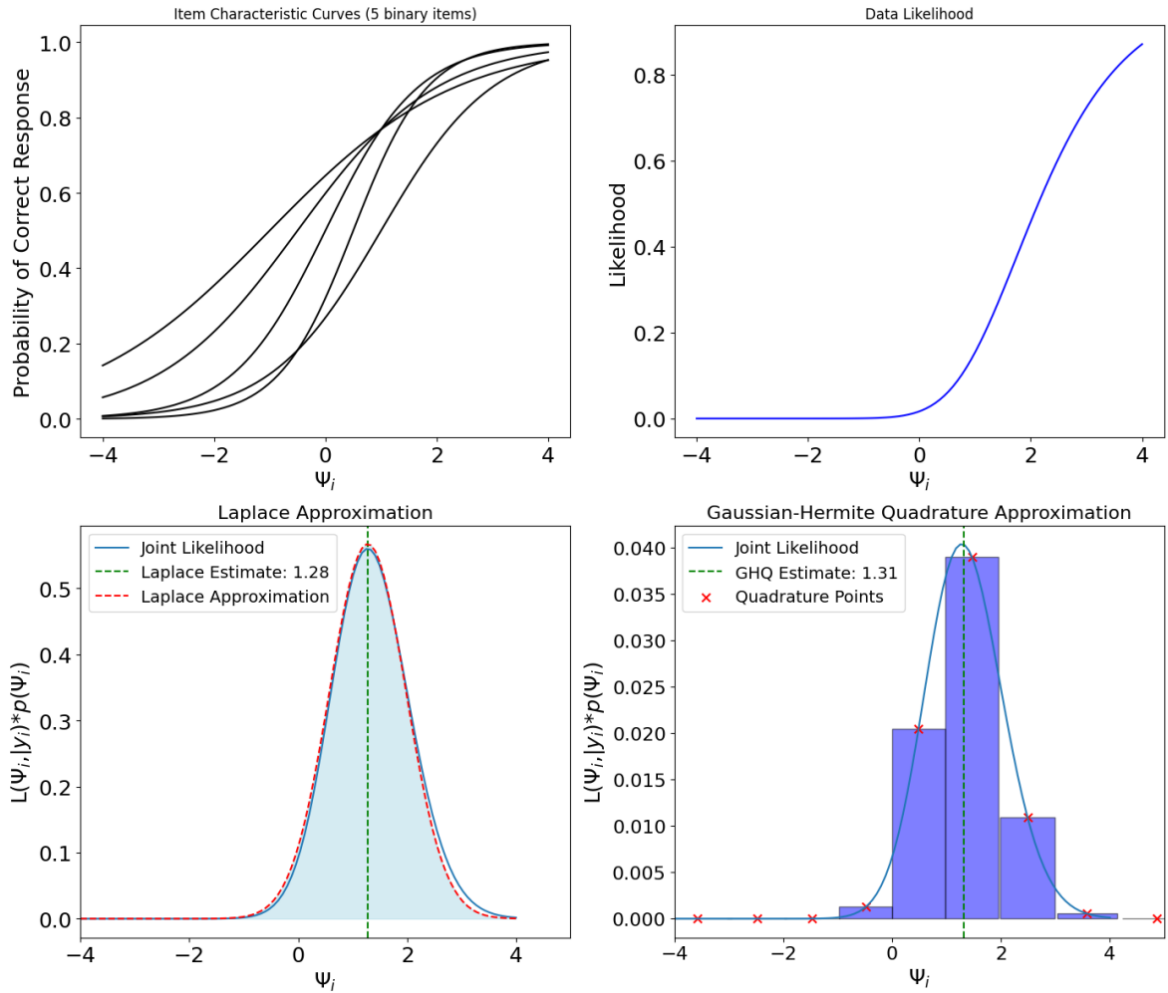

**Figure 1** Illustration of the Laplace and GHQ-EM likelihood approximation: item characteristic curves (top left), data likelihood of response pattern of 1,1,1,1,1 (top right), Laplace approximation to the joint-likelihood (bottom left), and GHQ-EM approximation to the joint likelihood (bottom right).

## Methods

To assess and compare the performance of Laplace and GHQ-EM a simulation study evaluating different scenarios ($s$) with one observation per subject (i.e., at baseline) was performed. Four sample sizes ($N_s$=50,100, 250,500) and two assessment lengths ($M_s = 5, 20$) were evaluated. The general workflow is described in Figure 2.

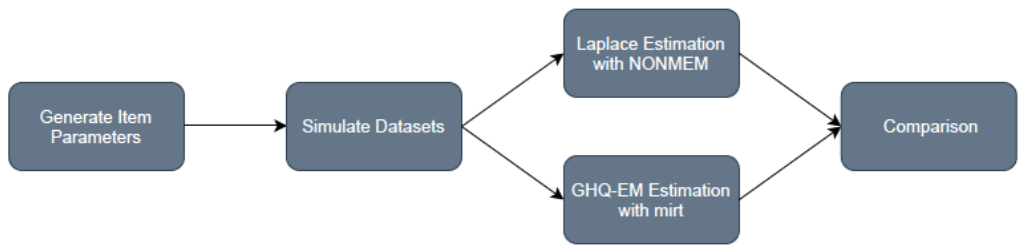

**Figure 2** Workflow

## Data generation

Ordered categorical items were simulated with 5 categories of responses (0-4). The item parameters were randomly sampled from a log-normal distribution (meanlog=0.05, sdlog=0.5) for the discrimination parameter or uniform distribution increasing across thresholds from the lowest threshold b1 (-2.5 to -1.1), b2 (-1 to -0.1), b3(0.1 to1) to the highest threshold b4 (1.1 to 2.5). (Figure 3).

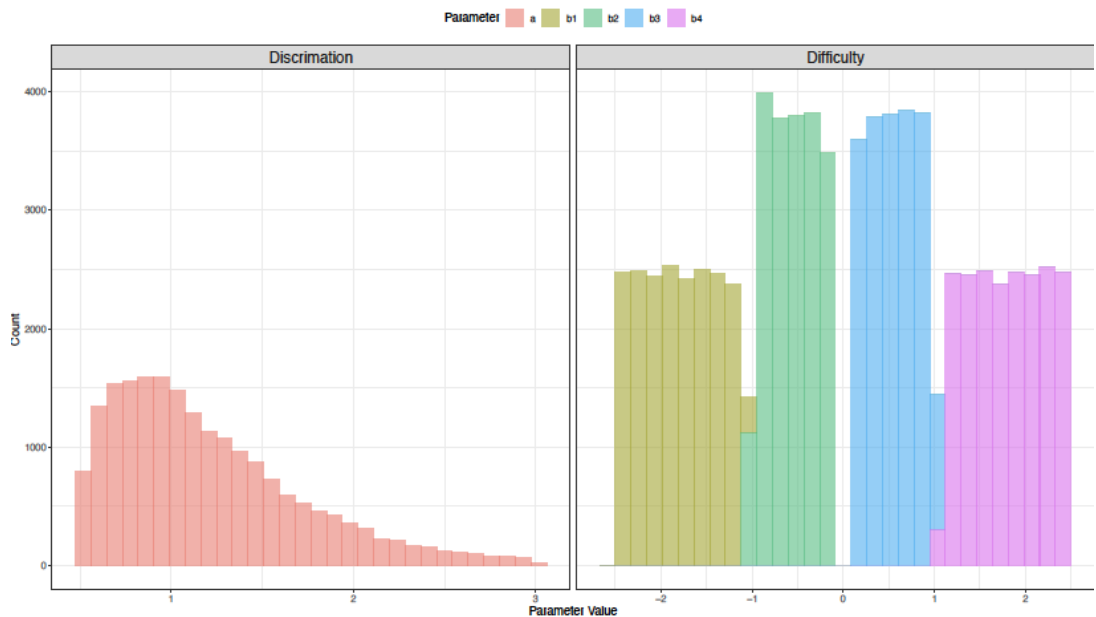

**Figure 3** Distribution of simulated item discrimination and threshold parameters

Replicate (R=1000) datasets were simulated for each sample size and item scenario using *mirt* from the set of simulated item parameters. Simulations of response categories were repeated until all items in each dataset contained all possible response categories.

## Graded Response model

A unidimensional graded response IRT model served as both the simulation and estimation model.

For subject $i$ and item $j$, the graded response model describes the probability of achieving a score of at least s as

$$P(Y_{ij} \geq s) = \frac{e^{a_j(\Psi_i - b_{j,s})}}{1 + e^{a_j(\Psi_i - b_{j,s})}}$$

and, consequentially, to achieve a score of exactly $s$ as

$$P(Y_{ij} = s) = P(Y_{ij} \geq s) - P(Y_{ij} \geq s + 1)$$

where $\Psi_i$ is the subject's latent variable value at baseline, $a_j$ is the item-specific discrimination parameter and $b_{j,s}$ is the threshold parameter for that specific item and score.

## Estimation methods

The R package *PIRAID* was used to autogenerate NONMEM model control files based on each dataset [10]. The initial estimates in NONMEM were set to the starting estimates obtained with *mirt* given the dataset using the item correlation matrix, to achieve similar starting conditions in both software. The upper bounds for the threshold fixed effects were set to 10, to set theoretically reasonable bounds without imposing tight restrictions in order to not give an advantage to NONMEM estimation. The latent variable was modeled through subject-specific random effect, assuming normal distribution with a mean of zero and fixed variance.

All parameters were jointly estimated using the Laplacian estimation method in NONMEM version 7.4.3 facilitated by PsN 4.7.15. Model estimation in *mirt* v1.31 was performed using stochastic EM with fixed quadrature [5].

All data handling and summarization was completed in R 3.5.2. Final parameter estimates from both *mirt* and NONMEM were converted to traditional IRT parameterization for results reporting.

## Assessment of performance

Item parameter recovery in relation to true parameters was evaluated for each item $j$, scenario $s$, and replicate $r$ in terms of estimation error ($e$), bias, and root mean square error (RMSE) which were defined as follows:

$$e_{j,s,r} = \hat{\theta}_{j,s,r} - \theta_{j,s,r}$$

$$Bias_{j,s} = \frac{1}{R}\sum_{r=1}^{R} e_{j,s,r}$$

$$RMSE_{j,s} = \sqrt{\frac{1}{R}\sum_{r=1}^{R} e_{j,s,r}^2}$$

where $\hat{\theta}_{j,s,r}$ is the estimated and $\theta_{j,s,r}$ is the true item parameter. In addition, a robust version of the RMSE (rRMSE) was evaluated, which removed the 1% most extreme estimation errors from both sides of the distribution.

On top of the parameter level comparison, the two algorithms were also compared in terms of the log-likelihood as reported in the respective software (i.e., OFV/2 for Laplace in NONMEM), run time and completion rate. Completion rate was determined by the number of successful models indicated by successful convergence in *mirt* and minimization successful in NONMEM out of total replicates.

Finally, as a measure of the resulting bias and precision on the total score levels, the expected total score (TS) was calculated as the sum of the probabilities of endorsing each response category as a function of ability ($\Psi$) across all items in a test, i.e.,

$$TS_{s,r} = \sum_{j=1}^{M_s} P(Y_{ij} = 1) + P(Y_{ij} = 2) \cdot 2 + P(Y_{ijk} = 3) \cdot 3 + P(Y_{ijk} = 4) \cdot 4$$

# Results

## Accuracy and Precision

The estimation error for each item parameter is presented in Table 1 for a sample size of 100 subjects. Tables summarizing sample size of 50, 250 and 500 are presented in Supplementary Online Resource 1. Initially the sample size of 50 was evaluated as a lower threshold to investigate worst-case scenario however the small sample size resulted in unreliable parameter estimation which did not add value to the overall estimation algorithm comparison. Therefore, results for N=50 are not shown for most of the comparison metrics. Overall *mirt* and NONMEM performed similarly with relatively low estimation error (median near zero). Precision increased as the number of subjects and items increased, indicating reasonable item parameter recovery.

**Table 1** Summary statistics of item parameter estimation error for sample size of 100 (items=20 and 5)

|  | N100 5 items | | | N100 20 items | |
|---|---|---|---|---|---|
|  | GHQ-EM | Laplace | | GHQ-EM | Laplace |
|  | R=1000 | R=1000 | | R=1000 | R=1000 |
| **a1** | | | | | |
| **Mean (SD)** | 0.0866 (0.536) | 1.75 (89.3) | | 0.0326 (0.293) | 0.0325 (0.292) |
| **Median [Min, Max]** | 0.0186 [-1.47, 8.65] | -0.00519 [-2.04, 6310] | | 0.0122 [-1.25, 2.32] | 0.0122 [-1.25, 2.25] |
| **b1** | | | | | |
| **Mean (SD)** | -0.129 (2.73) | -0.279 (1.23) | | -0.112 (1.72) | -0.122 (0.720) |
| **Median [Min, Max]** | 0.0137 [-43.2, 152] | -0.0236 [-12.8, 1.47] | | -0.00688 [-65.7, 156] | -0.00713 [-16.4, 1.51] |
| **b2** | | | | | |
| **Mean (SD)** | -0.0233 (1.05) | 0.0727 (0.519) | | -0.0329 (0.752) | -0.0359 (0.375) |
| **Median [Min, Max]** | 0.0108 [-10.7, 62.0] | -0.00267 [-10.6, 3.04] | | -0.00152 [-38.6, 70.9] | -0.00208 [-10.1, 2.85] |
| **b3** | | | | | |
| **Mean (SD)** | 0.0348 (1.34) | 0.0915 (0.539) | | 0.0285 (0.556) | 0.0298 (0.365) |
| **Median [Min, Max]** | 0.00458 [-80.7, 10.8] | 0.0175 [-6.12, 9.63] | | 0.00470 [-38.1, 18.3] | -0.00460 [-1.89, 7.93] |
| **b4** | | | | | |
| **Mean (SD)** | 0.123 (3.65) | 0.299 (1.26) | | 0.111 (1.40) | 0.118 (0.724) |
| **Median [Min, Max]** | 0.00670 [-221, 34.9] | 0.0419 [-2.95, 14.6] | | 0.000897 [-104, 39.6] | 0.000522 [-1.41, 15.8] |

The RMSEs calculated for all scenarios and item parameters are displayed in Figure 4. While Figure 4a shows the standard RMSE, 4b focuses on the more robust RMSE where the 2% most extreme values have been removed before the calculation. The standard RMSE reveals a differential image between discrimination and threshold parameters. While GHQ-EM was able to estimate the discrimination parameters with a much lower RMSE, Laplace generally performed better for the threshold parameters in the presence of fewer items or subjects. The results are more uniform when considering the rRMSE, with GHQ-EM consistently showing lower or equal rRMSE values for all scenarios and parameters.

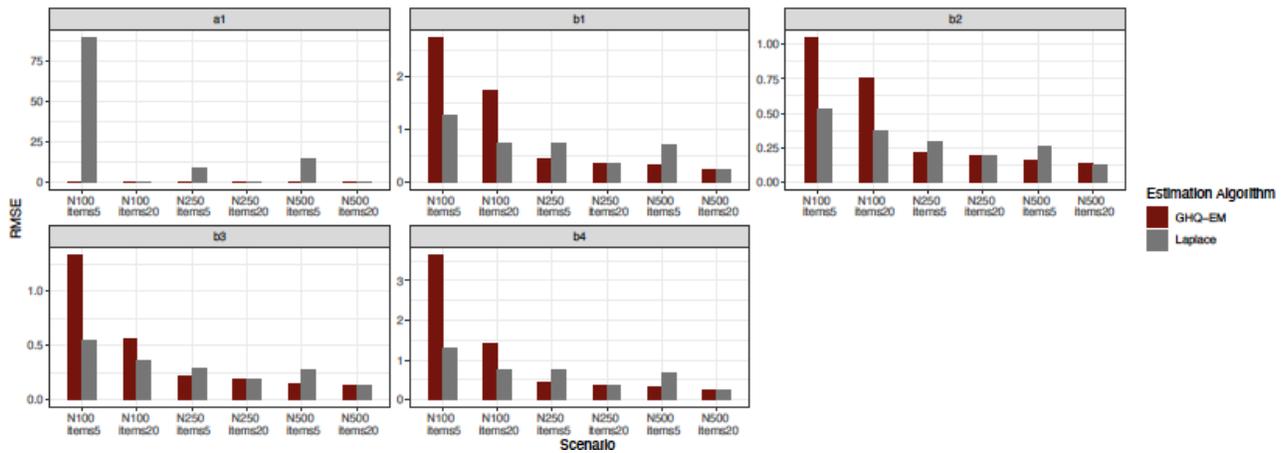

**Figure 4a** Item parameter RMSE for each sample size and item scenario

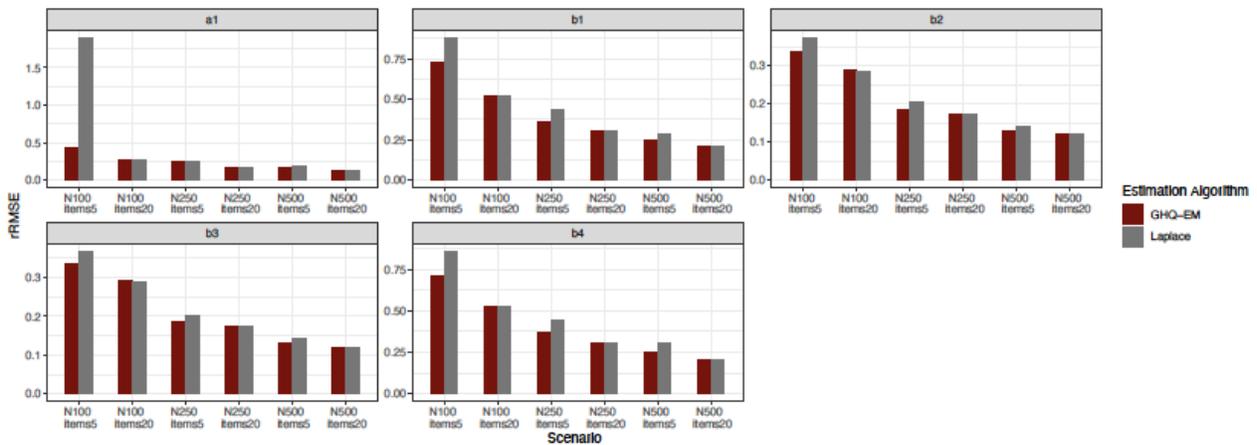

**Figure 4b** Item parameter rRMSE for each sample size and item scenario

### Agreement between algorithms

The agreement in item parameter estimates between algorithms across all scenarios can be appreciated in Figure 5. Points that appear on the line of identity intend to indicate perfect agreement between both algorithms. For the scenario with 5 items and 100 subjects, the data points scatter around the line of identity with approximate agreement between both algorithms for most samples but also with a considerable number of samples with large differences. With an increasing number of data points (i.e., more items and subjects) the number of samples with a disagreement in estimates decreases. From the scenario with 20 items and 250 subjects, perfect agreement between both algorithms is reached for all samples.

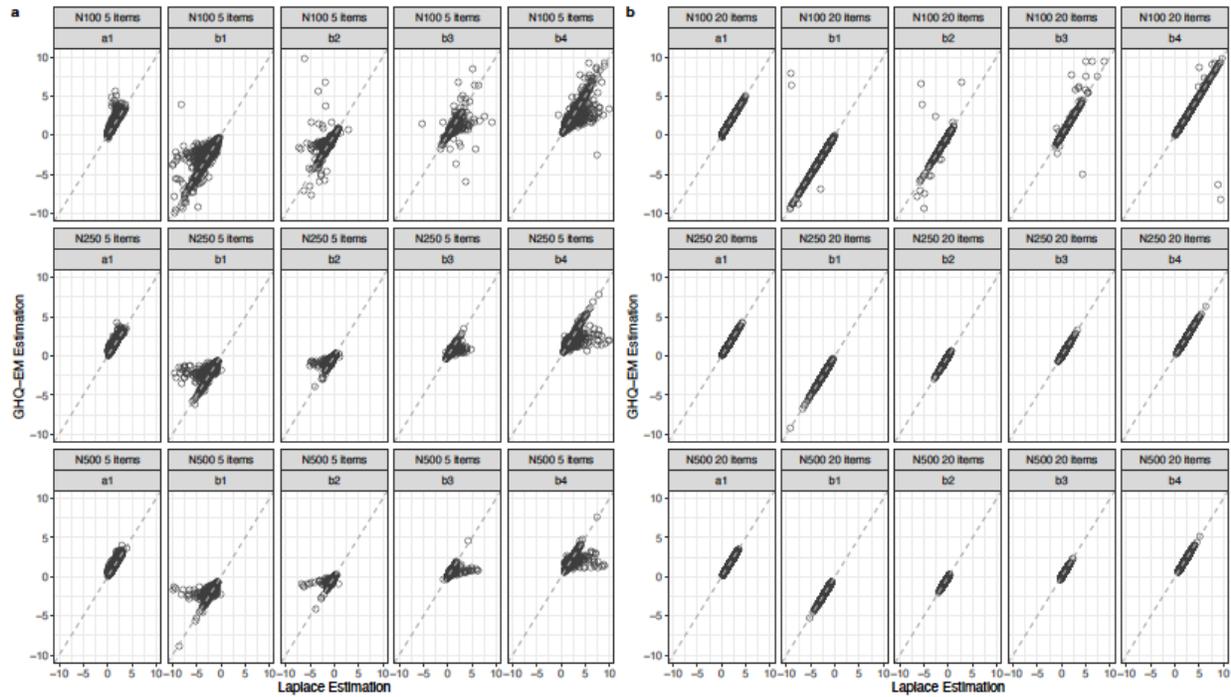

**Figure 5** Item parameter estimates   a) 5 items b) 20 items

The log-likelihood comparison between estimation algorithms for 5 and 20 items are presented in Figure 6. Laplace and GHQ-EM performed equally well for the scenarios with 20 items, which presents more observations per subject, regardless of sample size. In the scenarios with 5 items Laplace log-likelihood was lower in 2.7%, 4% and 8% of the cases for 500, 250 and 100 subjects respectively. However, the difference between estimation values was less than 5%.
It is possible that local minimas may have been found in these cases.

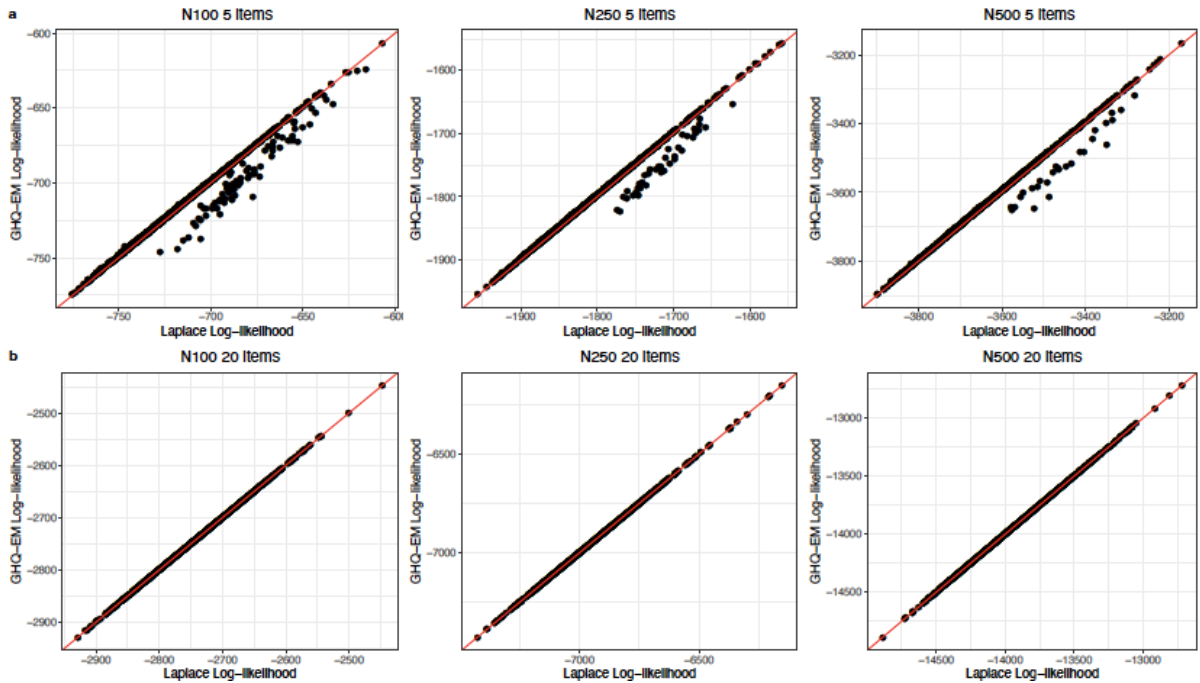

**Figure 6** Model log-likelihood for single latent variable scenario a) 5 items b) 20 items

## Estimation error on the expected total score level

The final item parameters from both algorithms as well as the true item parameters were used to calculate the expected total score for a latent variable range from -4 to 4 for all scenarios and replicates. The resulting estimation error on the expected score level normalized by the number of items is represented in Figure 7. The differences in item parameter estimates, as appreciable from Figure 6, are greatly reduced when translating results to the total score scale. Differences between algorithms are only visible for the 5 item scenarios at the tails of the latent variable distribution (PSI values smaller than -2 or larger than 2). Generally, GHQ-EM showed slightly lower bias and higher precision than Laplace.

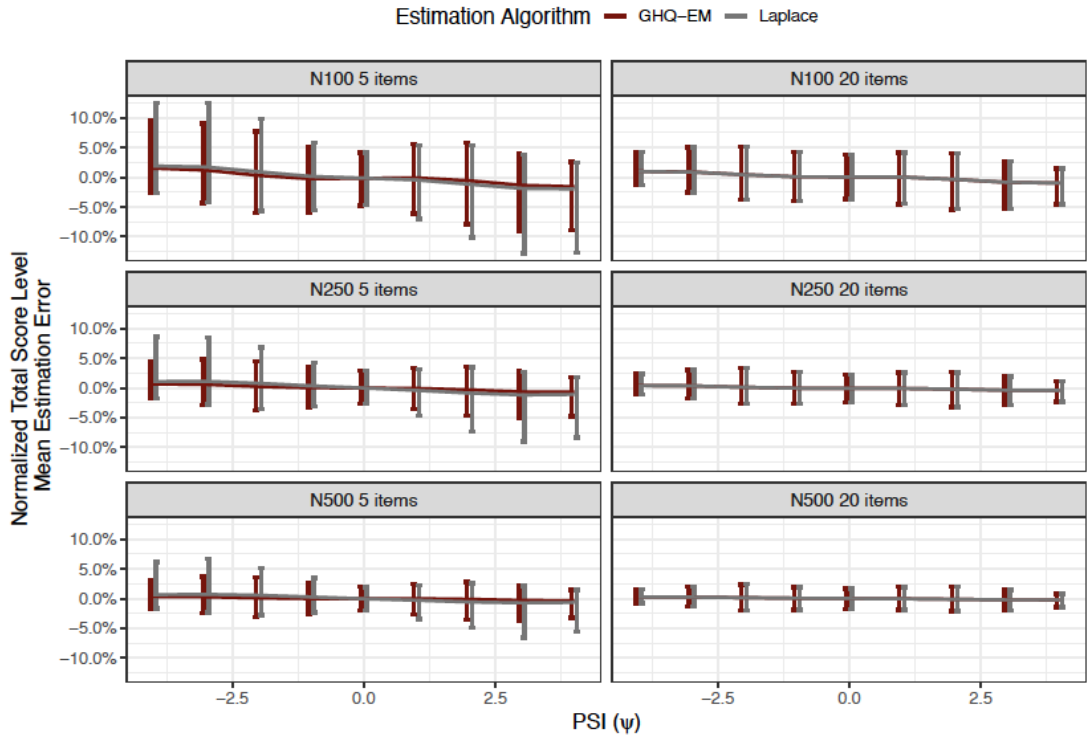

**Figure 7** Mean estimation error (2.5% and 97.5% percentile) in expected total score normalized by the number of items.

## Completion Rate and Run time

The algorithm comparison of completion rate is presented in Table 2. Consistently across all scenarios GHQ-EM has a completion rate of 98% or higher. With Laplace the completion rate ranges from approximately 84% to 99 %; increasing with an increase in sample size or addition of items. Computationally, GHQ-EM is many orders of magnitude faster than Laplace with an average model estimation time of approximately 19s seconds compared to approximately 5 minutes. In a real data scenario where models may become more complicated or data is less well behaved it is expected that the run time will increase considerably. However, the increased speed observed when using a fixed grid compared to a search algorithm will still be substantial.

**Table 2** Model completion rate comparison

| Scenario | Completion Rate | |
| --- | --- | --- |
| | mirt | NONMEM |
| N50 5 items | 98.9% | 84.3% |
| N50 20 items | 99.3% | 93.0% |
| N100 5 items | 98.2% | 91.0% |
| N100 20 items | 99.9% | 97.0% |
| N500 5 items | 99.6% | 96.5% |
| N500 20 items | 99.9% | 98.9% |

# Discussion

In this work, we compared two estimation algorithms for the item parameter estimation in IRT models with a pharmacometric focus; GHQ-EM employed in mirt and Laplace employed in NONMEM. Each estimation algorithm presents beneficial features as well as limitations that could impact parameter estimation in an IRT framework.

The results of our investigation show that in terms of item parameter recovery in scenarios with at least 20 items and more than 100 subjects, parameters were estimated with good precision and were essentially identical between estimation algorithms. These observed differences in bias and precision increased for smaller sample and item sizes, however the magnitude of this finding is dependent upon the evaluation metric applied (i.e., RMSE vs rRMSE). Lastly, GHQ-EM algorithm exhibited significantly faster runtimes compared with Laplace. This presents an opportunity to leverage *mirt* to provide item parameter estimates for a pharmacometric IRT analysis using NONMEM; among other potential applications.

We used two levels of RMSE as seen in figure 5, rRMSE was used to represent the summary of the robust central tendency with the upper and lower 1% of the data removed. This removal equated to approximately 10 records. In our case when there appeared to be differences in bias or precision which indicated a more superior performance of one estimation algorithm over another at the RMSE level, utilizing the rRMSE metric these differences were now negligible. The decision whether to use RMSE vs rRMSE metric depends on the methodological question proposed or type of analysis. RMSE is a more global measure which includes outliers and could be useful, for example in a simulation study where understanding of outliers is important. Alternatively in a case where there is a single trial analysis the modeler will be able to visually identify any outliers in parameter estimates when non-plausible values appear and can address accordingly during analysis.

One factor that influences estimation and subsequently impacts RMSE calculation is the parameter boundary setting. In our initial exploration in NONMEM, the upper bounds for THETA of the threshold parameters were set to 50 for b2, b3 and b4. Not surprisingly when the upper bounds were set to 50, creating a wider parameter search space, this resulted in less optimal parameter estimation. Ultimately, for subsequent evaluations the upper bounds were set to 10, which presents a more reasonable parameter boundary within expectation of the scale. In *mirt* we did not employ user defined boundaries for the parameters and accepted the default boundaries of negative to positive infinity. The accuracy still observed in *mirt* appears to indicate that GHQ-EM is less impacted by the boundaries. The implementation of bounds analogous to those set in NONMEM is not straightforward considering the different parameterization in *mirt* (i.e., slope- intercept).

Our explanation for these findings is rooted in the fundamental approach of estimation within each algorithm. In Laplace estimation requires taking the second derivative with respect to each individual ETA therefore the search for the mode of distribution is time consuming and potentially less robust under certain conditions. In contrast, evaluation with a fixed grid, as was done in GHQ-EM is fast and reliable. Furthermore, in the case of Laplace one could expect that estimates could be impacted by shrinkage. This could be a possible explanation for the lower RMSE for the discrimination (a) parameter in GHQ-EM compared with Laplace.

In this work for the first time the impact of item parameter precision on expected total score was evaluated. In the clinical trial setting the endpoints based on clinical rating scale assessments are evaluated at the total score level therefore often modeling results are often communicated on the same scale. While there were minor differences between estimation algorithms and some imprecisions in the item parameter estimates this did not translate to a meaningful difference in the resulting expected score. However, one could also argue that the imprecisions at the item level mostly affect the tails of the latent variable distribution; with mean bias in total expected score increasing as the latent variable moves from the population mean. It is important to note that the reported expected total score in this work does not take into account the weighting at the item level in terms of parameter uncertainty, therefore if the weighting is less precise or the model is misspecified the impact on total score would be greater.

In practice there may be a desire to analyze smaller datasets with an IRT model. Testing this more extreme case revealed that 50 subjects with a single time point, may be too few subjects for reliable item parameter estimation. We found that the parameter estimate agreement between estimation algorithms was less than ideal, with many estimates being close to the upper and lower ends of the -10 to 10 range. This was also visible in the range of bias measurements for each algorithm (Supplemental online resource 1). In the presence of longitudinal data, the sample size would now be N_timepoints*N_subjects (e.g., 50 subjects at 5 time points = 250), this illustrates that indeed scenario of 50 is very small and less relevant.

One limitation of this work is that in our investigation we elected to evaluate item parameter estimation under the pretense of developing ICCS using a single time point. This was done in order to simplify and limit the number of scenarios, although in pharmacometrics applications there are generally many timepoints.

It is worth noting that there are alternative estimation algorithms in NONMEM and *mirt*. The *mirt* package in general is recommended for more complex models (e.g., multiple latent variables) and the additional algorithms are not expected to present additional benefit for single latent variable models, therefore these additional algorithms were not evaluated here.
In NONMEM there is also functionality to execute EM methods that are capable of handling categorical or discrete data. Importance sampling and stochastic approximation expectation maximization (SAEM) could potentially be interesting approaches from a theoretical point of view because estimation is less impacted by shrinkage. However, our pilot study showed a high failure rate resulting in non-plausible estimates therefore these were not explored further. One reason for this is, in general, IRT models contain relatively low number of random effects and these methods tend to perform well when parameterized with many random effects [1]. Despite our initial findings there may still be value trying these methodologies in practice.

One possible future extension of this work is the inclusion of longitudinal data. Incorporation of longitudinal data in NONMEM is more straight forward and much more flexible than within *mirt*. The implications of how the longitudinal data is incorporated in the IRT model, regarding development of the ICCs, could potentially impact the overall estimation precision and has yet to be explored.

## Conclusion

Item parameter recovery and overall model fit were similar between Laplace and GHQ-EM. The ease of use, speed of estimation and relative accuracy of the GHQ-EM method employed in the existing *mirt* R package make it a well-suited alternative or supportive analytical approach to NONMEM for potential IRT applications.

**Acknowledgements**

The authors would like to thank Mats O. Karlsson for his input during the research. Open access funding provided by Uppsala University. This study was supported by the Swedish Research Council Grant 2018-03317.

**Disclosures**

L. Arrington and S. Ueckert were affiliated with Uppsala University at time this work was performed. L. Arrington is currently employed by Amgen Inc. but still affiliated with Uppsala University. S. Ueckert is currently employed by Ribocure Pharmaceuticals AB.